\documentclass[%
 aip,
 amsmath,amssymb,
 reprint,%
]{revtex4-1}

\usepackage{graphicx}
\usepackage{dcolumn}
\usepackage{bm}

\usepackage[utf8]{inputenc}
\usepackage[T1]{fontenc}
\usepackage{mathptmx}
\usepackage{etoolbox}
\usepackage{tikz}
\makeatletter
\def\@email#1#2{%
 \endgroup
 \patchcmd{\titleblock@produce}
  {\frontmatter@RRAPformat}
  {\frontmatter@RRAPformat{\produce@RRAP{*#1\href{mailto:#2}{#2}}}\frontmatter@RRAPformat}
  {}{}
}%
\makeatother
\begin{document}

\preprint{AIP/123-QED}

\title{In-Depth Investigation of Phase Transition Phenomena in Network Models Derived from Lattice Models}
\author{Yonglong Ding}
\affiliation{School of Physics, Beijing Institute of Technology, Beijing 100081, China}
\affiliation{Beijing Computational Science Research Center, Beijing 100193, China}
 \email{ylding@csrc.ac.cn.}


\date{\today}

\begin{abstract}
Lattice models exhibit significant potential in investigating phase transitions, yet they encounter numerous computational challenges. To address these issues, this study introduces a Monte Carlo-based approach that transforms lattice models into a network model with intricate inter-node correlations. This framework enables a profound analysis of Ising, JQ, and XY models. By decomposing the network into a maximum entropy and a conservative component, under the constraint of detailed balance, this work derive an estimation formula for the temperature-dependent magnetic induction in Ising models. Notably, the critical exponent $\beta$ in the Ising model aligns well with established results, and the predicted phase transition point in the three-dimensional Ising model exhibits a mere $0.7 \%$ deviation from numerical simulations.
\end{abstract}

\maketitle
\section{Introduction}

In the realm of many-body physics, direct computational methods often face an exponential increase in complexity as the number of particles grows, swiftly surpassing the computational capabilities of current hardware\cite{ortega2015fpga,yang2019high,preis2009gpu,meredith2009accuracy}. To delve into the intricacies of many-body phenomena, lattice models find extensive application, with Monte Carlo methods\cite{RevModPhys.73.33,sarrut2021advanced,luo2022hybrid} proving notably efficient for various lattice models. The evolution of Monte Carlo\cite{tang2018role,carlson2015quantum,lynn2019quantum,nemeth2021stochastic,rioux2022monte,andersen2019practical}, spanning many years since its inception, has established its prominence in physics, particularly in computations related to phase transitions, ground-state energy and wavefunction exploration, and unraveling the mechanisms underlying phenomena like superconductivity\cite{dai2023quantum,huang2019antiferromagnetically,kato2020many}. Graph theory, an age-old mathematical pursuit illustrated by classics such as the Seven Bridges of Königsberg problem, continues to play a pivotal role. The application of graph theoretical methods demonstrates substantial prowess in addressing a spectrum of computational challenges.

In addressing diverse physical phenomena, a multitude of Monte Carlo computational techniques has emerged, encompassing methodologies such as projector quantum Monte Carlo\cite{schwarz2017projector} Continuous-time quantum Monte Carlo\cite{rubtsov2005continuous} diagrammatic Monte Carlo\cite{prokof2008fermi} and determinant Monte Carlo\cite{moutenet2018determinant}. Nonetheless, Monte Carlo methods exhibit certain constraints. Notably, in Fermi systems, challenges arise, exemplified by the sign problem\cite{henelius2000sign,alexandru2022complex}, prompting extensive discourse and scrutiny within the physics community\cite{mondaini2022quantum,zhang2022fermion,berger2021complex}. Furthermore, graph theory often intersects with NP problems. Despite the substantial simplification of relationships between nodes in graph computation\cite{sanchez2020learning,schuetz2022combinatorial}, the associated problem instances are frequently complex, yielding noteworthy research outcomes. Various methodologies have been explored to navigate challenges inherent in graph computation. Quantum computing, in its current trajectory, appears to present a fresh perspective for approaching problems within the domain of graph computation.

In this study, I introduce a novel algorithm built upon the foundation of the lattice model's Monte Carlo algorithm. This innovative approach involves the transformation of the lattice model into a network model. Through the employment of network models, this work aim to capture the intricate statistical patterns inherent in lattice models, patterns that are challenging to address or yet to be uncovered through conventional methods. Given the Hamiltonian of a lattice model, this paper categorize all potential lattice points based on their interactions. Subsequently, these distinct categories of lattice points are transformed into nodes within the network model. By analyzing the updating dynamics of the network model through Monte Carlo simulations, this work integrate principles of maximum entropy. This approach allows us to establish a correspondence between phase transitions in lattice models and distinct network structures within the network model. To substantiate the validity of this methodology, this paper apply this network computing model to calculate thermodynamic phase transitions in the Ising model, encompassing both trivial and non-trivial quantum phase transitions like the XY model\cite{barouch1970statistical,kosterlitz1974critical}, JQ model\cite{desai2020first}. The results obtained align with our expectations. Additionally, we engage in a comprehensive discussion on how the renormalization group\cite{dupuis2021nonperturbative,eichhorn2023microscopic,soejima2020efficient,balog2019convergence,zhai2021low,ganahl2023density} is manifested within the realm of network computation.

\section{Theory}

Herein, section A establish the conversion relationship between lattice points in lattice models and nodes in network models. Subsequently, section B elaborate on the construction and decomposition of the network model. Ultimately, section C discuss the implications of this network model for thermodynamic and quantum phase transitions.

\subsection{network node}

Lattice models play a pivotal role in diverse applications within the realm of physics. In the context of these models, this approach commences with a specified Hamiltonian, intricately categorizing each lattice point based on its unique attributes and the extent of interactions with its nearest neighbors. These categorized lattice points correspond to nodes in graph theory, with the size of each node proportionate to the ratio of its associated lattice point's category to the total number of points.

This paper illustrate the concrete example of converting lattice points into network nodes:
The transformation involves converting the lattice point model into a multi-layer network model. In the absence of external fields, lattice points sharing the same energy are grouped into the same layer within the network structure. Within each layer, lattice points exhibiting identical orientations are assigned to the same network node. The numerical representation of these nodes corresponds to the weight attributed to such lattice points. I provide specific conversion instances from the perspectives of thermodynamic phase transitions, trivial quantum phase transitions, and non-trivial quantum phase transitions.

In this endeavor, I embark on elucidating the interplay between lattice models and graph computation models, commencing with the paradigmatic example of the two-dimensional Ising model. This article takes the ferromagnetic case as an example.
\begin{equation}
    H=J\sum_{<ij>}S_{i}\cdot S_{j},S_{i}=\pm1,(J<0)
\end{equation}
In the context of the Ising model, each lattice point manifests two conceivable spin states—up or down—while only factoring in nearest-neighbor interactions. Predicated upon the magnitudes of these interactions, all lattice points are systematically categorized into 5 distinct classes. For those lattice points exhibiting an up spin, a nuanced classification ensues, stratified into 5 subclasses contingent on the tally of neighboring points with identical spins, inclusive of the point itself. Analogously, spin-down lattice points undergo a parallel classification, culminating in a total of 10 unique categories. Utilizing the aforementioned classification, the Hamiltonian can be meticulously reorganized into the subsequent mathematical expression:
\begin{equation}
    H=J\sum_{k}\sum_{<ij>}S_{i}\cdot S_{j},S_{i}=\pm1, i=1:n, j=1:4, k=1:10
\end{equation}

\begin{figure}
\centering
\begin{tikzpicture}

\scope[nodes={inner sep=0,outer sep=0}]
\node[anchor= east] (a)
  {\includegraphics[width=4cm]{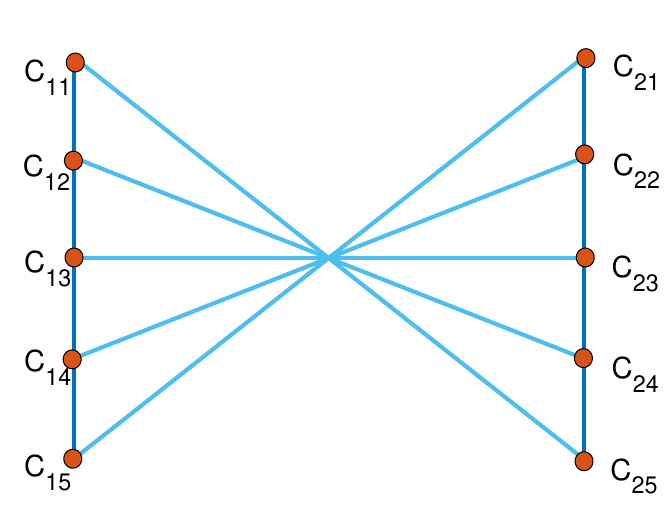}};
\node[anchor= west] (b)
  {\includegraphics[width=4cm]{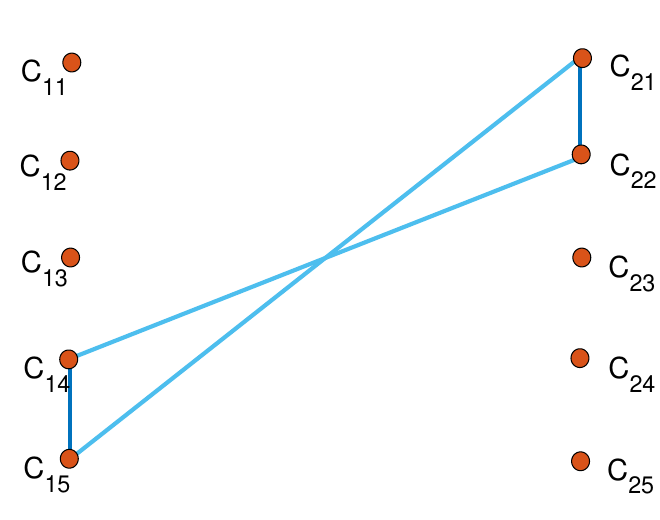}};
\endscope
\foreach \n in {a,b} {
  \node[anchor=west] at (\n.north west) {(\n)};
}

\end{tikzpicture}

\caption{\label{fig:1} The red dots symbolize the network nodes, each corresponding to a distinct type of lattice point within the two-dimensional Ising model. Meanwhile, the blue bonds illustrate the transformational relationships among these various nodes. (a) (b) represents two distinct and stable network configurations observed within the Ising model. c represents the weight of the node. }
	\label{fig2}
 
\end{figure}

By regarding these 10 categories as nodes, the value assigned to each node encapsulates the prevalence of its corresponding category. This transformative process effectively transmutes the lattice model of an infinitely extended two-dimensional square grid into a graph endowed with ten nodes, intricately interlinked in accordance with the underlying classification schema.

Distinct types of lattice models exhibit unique functional mechanisms, enabling their transformation into network models, analogous to the two-dimensional Ising model, through Monte Carlo updating operations and the intricate interplay of interactions. In the realm of the Ising model, when deploying the Monte Carlo algorithm for computational pursuits, the act of flipping a solitary lattice point reverberates to impact its four proximal neighbors. For a given lattice point, I adeptly discern all conceivable categories and their attendant probabilities for the contiguous four points through a probabilistic framework. Within the graph model paradigm, this translates to a scenario where, in the graph comprising 10 nodes, a maximum of 5 nodes can undergo changes simultaneously. As illustrated in Fig.~\ref{fig:1}(a). 

In the context of ferromagnetic systems, during the process of transformation of various nodes, lattice point $c_{15}$, characterized by identical nearest neighboring lattice points as itself and maintaining the same spin direction as the initial moment, is designated as an internal lattice point in this study. Conversely, lattice point $c_{14}$, possessing only one different type of nearest neighbor lattice point while retaining the spin direction consistent with the initial state, is categorized as a boundary lattice point. Additionally, $c_{13}$ and $c_{23}$, exhibiting identical nearest neighbor lattice points of two distinct types, are labeled as central lattice points.

In the realm of higher-dimensional Ising models, a parallel pattern emerges. For instance, a three-dimensional Ising model manifests into 14 distinct categories, while its four-dimensional counterpart expands to 18 categories. Extending this logic, an n-dimensional Ising model can be systematically categorized into $4n+2$ classes, thereby seamlessly translating into a graph computational model boasting $4n+2$ nodes.

In our investigation,This article takes the case of the JQ model as an example. I address the paradigm of a trivial quantum phase transition model, introducing a competitive term denoted as g. We commence with the simplest model, employing the classification method elucidated earlier. For both spin-up and spin-down lattice points, there exist 8 categories in their vicinity, resulting in a total of 16 classes. It is important to note that the classification scheme may vary for different quantum phase transition models. Subsequent to these initial steps, I systematically transform the model into a graph computational representation for simulation purposes. 
\begin{equation}
    H(g)=H_{0}+gH_{1}
\end{equation}

\begin{equation}
    [H_{0},H_{1}]\ne 0
\end{equation}




During the process of varying $g$, the occurrence of the trivial quantum phase transition is manifestly evident, as it is characterized by a change in the node possessing the lowest energy.

Transitioning to the realm of non-trivial quantum phase transitions, I specifically explore the two-dimensional $XY$ model within the domain of topological phase transition models. The $XY$ model introduces complexity, as lattice points no longer exhibit only two possibilities $(up and down)$, but rather an infinite range of potential states. Consequently, our previous classification methodology becomes impractical. To address this, I implement a simplification by initiating a preliminary classification based on clockwise or counterclockwise rotations, assigning m possible values to each lattice point. As m approaches infinity, the model converges to the $XY$ model. With this approach, I classify the system by determining, for a given value of r, the categories for each lattice point surrounded by four others . This results in a count of $C_{r}^1+3C_{r}^2+3C_{r}^3+C_{r}^4$  categories. Thus, the two-dimensional infinite lattice model transforms into $rC_{r}^1+3rC_{r}^2+3rC_{r}^3+rC_{r}^4$ nodes  in the graph computational representation.






\subsection{network model}

In the realm of computational modeling, the Monte Carlo algorithm serves as a pivotal technique, with importance sampling as one of its fundamental components. When examining models exhibiting nearest-neighbor interactions, such as the Ising model and Potts model, the Hamiltonian, which characterizes these interactions, is represented through bonds connecting adjacent lattice points. Specifically, in lattice-based models, the interaction between any two neighboring lattice points is quantified by the magnitude of the bond connecting them.

In the context of the Ising model, transformations resulting from bond changes yield four potential scenarios. However, for the sake of simplicity and computational efficiency, we focus primarily on two distinct bond values, abstracting away from the intricate details. The ratio of these two bond types, which holds significant implications for thermodynamic phase transitions, can be approximated through meticulous calculations based on the principle of detailed balance.

The flipping of a lattice point, a fundamental operation in our simulations, involves alterations in the bond configurations. These changes correspond to variations in the quantities of different bond types, reflecting the dynamic nature of the system. For instance, as the temperature increases, the relative abundance of bonds with higher energy levels tends to increase, reflecting the thermodynamic behavior of the system.

Based on my previous classifications, nodes in the Ising model possess varying quantities of positive and negative bonds. At a given temperature, the Ising model attains equilibrium, resulting in a constant energy value derived from the Hamiltonian. Therefore, the ratio of these bonds attains a specific equilibrium value. In a scenario where all lattice points are in a state of complete randomness, it follows the principle of maximum entropy. Under such conditions, the weights assigned to different types of lattice points can be derived directly from the ratio of positive to negative bonds, leveraging the principles of classical probability $C_{2n}^{l}b^{2n-l}(1-b)^l$. In this context, the symbol "l" denotes the kth layer of the network, where each layer embodies distinct magnitudes of nearest-neighbor interaction energy. The variable "n" signifies the dimensionality of the Ising model. For the two-dimensional Ising model, depicted in Fig.~\ref{fig:1}, b represents the weight of the bond with higher energy among all bonds. the following mathematical expression can be derived:

\begin{equation}
\label{eq:5}
   c_{11}+c_{12}=C_4^{0}b^4=b^4 
\end{equation} 
\begin{equation}  
\label{eq:6}
   c_{21}+c_{22}=C_4^{1}b^3(1-b)=4b^3(1-b) 
\end{equation} 
\begin{equation}
\label{eq:7}
   c_{31}+c_{32}=C_4^{2}b^2(1-b)^2=6b^2(1-b)^2 
\end{equation} 
\begin{equation}
\label{eq:8}
   c_{41}+c_{42}=C_4^{3}b(1-b)^3=4b(1-b)^3 
\end{equation} 
\begin{equation}
\label{eq:9}
   c_{51}+c_{52}=C_4^{4}(1-b)^4 =(1-b)^4 
\end{equation}

Furthermore, in the context of the XY model, lattice points of distinct types possess different energies. Transitions between lattice points of varying energies also adhere to detailed balance conditions. Thus, the prerequisites for applying the Monte Carlo algorithm are met. Unlike traditional many-body Monte Carlo algorithms, this solution doesn't directly apply the renormalization group algorithm. Nevertheless, implicit aspects of the renormalization group algorithm are present. In this work, the varying impact of each sampling on node values corresponds to the renormalization group algorithm in many-body computational simulations. Updates to lattice models across various scales directly correlate with the extent of change exhibited by individual nodes during the iterative update process of the corresponding network models.

Due to the infinite orientations of lattice points in the XY model, there correspond an infinite number of nodes and layers in the network model. For computational convenience, the orientations are fixed to a determined number, denoted as m, which directly determines the number of layers and nodes. Assuming that each step changes the orientation by $1/r$, resulting in either clockwise or counterclockwise rotation, the change in node weights affects only the nearest neighboring network nodes. Through this method, the XY model is approximately transformed into a multi-layer network structure.

\subsection{phase transition}

Prior discussions have established a direct mapping between the lattice model and its network representation. To further our understanding, in the context of the Ising model, we now explore two extremal scenarios from a spin-based perspective. Firstly, we consider a scenario where the lattice sites predominantly align in a unidirectional manner. Conversely, in the second scenario, we investigate a configuration where the distribution of lattice sites with different spin orientations is as uniform as possible. 

At extremely low temperatures, the model assumes a stable low-energy state. In the ferromagnetic context of the two-dimensional Ising model, when all spins are initially aligned, the model attains a stable configuration with the majority of spins oriented in the same direction. Post-phase transition, the model attains a state adhering to maximum entropy, exhibiting near-equal probabilities for spins pointing up or down. Consequently, the model's state can generally be expressed as a weighted sum of two extreme configurations: the most ordered and the most disordered states. In this context, the network structure under general conditions can be decomposed into a maximum entropy structure and a conservative structure, as illustrated in Fig.~\ref{fig:1}. Prior to the phase transition point, these two network structures compete, achieving a balanced state under varying conditions, consistent with the competition between free energy and entropy principles governing the Ising model's phase transition.

The network structure exhibiting maximum entropy, as depicted in Fig.~\ref{fig:1}(a), aligns the weights of each node with the probability distribution calculated directly using the bar methodology. Specifically, considering the two-dimensional Ising model as an example, the node weights adhere to the formulation presented in Eq.~\ref{eq:5}-Eq.~\ref{eq:9}. Notably, within the same layer of Fig.~\ref{fig:1}(a), the weights of the left and right nodes are equivalent, indicating that the weights of all network nodes can be precisely determined at a specified temperature. Furthermore, the weights of nodes representing spin-up and spin-down states are identical, suggesting that despite variations in node weights as temperature increases, the magnetic induction intensity remains zero.

A distinctive class of nodes within this structure is the centrally positioned node, as previously defined. Assuming an initial condition where the weight of the central node is unity and the weights of all other nodes are zero, it becomes evident that this maximum entropy configuration persists across various temperatures. The central node assumes a pivotal role analogous to a particle in a physical system. In contrast, the absence of such a central node, such as in the Ising model with an odd number of lattice points exhibiting nearest-neighbor interactions, would yield distinct properties and behaviors.

In the context of the conservative structure, as depicted in Fig.~\ref{fig:1}(b), exemplified by the two-dimensional Ising model, initially, spins are aligned predominantly in a single direction, with the weight of node $c_{15}$ approaching unity. Through Monte Carlo simulations, a fraction of lattice points undergoes flipping, inducing a shift in the weights of adjacent lattice points towards neighboring upper nodes. This phenomenon is effectively captured by the transformation of internal lattice points into boundary points, reflecting dynamic variations in the weights between boundary and internal nodes. Analogically, this process can be envisioned as a yarn ball unraveling, thinning, and elongating with rising temperatures, akin to the formation of fractal structures. The central node is omitted in this structure due to its inherent association with the maximum entropy configuration, which constitutes the foundation for subsequent computational investigations.

Next, considering the topological properties, in a lattice point model with nearest-neighbor interactions, such as the XY model, for a given square lattice, each lattice point interacts only with its four nearest neighboring lattice points. These interactions can be straightforwardly categorized into two pairs of interactions parallel and perpendicular to the vortices.
Given the weights of lattice points with different energies and the topological properties, the farther a lattice point is from the center of the vortex, the lower its energy. By selecting a radius from the center of the vortex, the total of parallel interactions on the same radius circle around the vortex center remains constant within a certain temperature range. However, the weights of lattice points with different energies change, causing variations in the perpendicular interactions to the vortex center.
As the temperature increases, the weights of lattice points with different energies change, with higher energy lattice points increasing in weight. Consequently, the perpendicular interactions to the vortex center also change. If the inner pair of interactions exceeds the outer pair, the overall spiral behaves like an attractive force; otherwise, it forms a repulsive force. The transition from attraction to repulsion signifies a topological phase transition. The above text provides a possible explanation for the KT phase transition in network models.

\section{Calculation}

\begin{figure}
\centering
\begin{tikzpicture}

\scope[nodes={inner sep=0,outer sep=0}]
\node[anchor= east] (a)
  {\includegraphics[width=4cm]{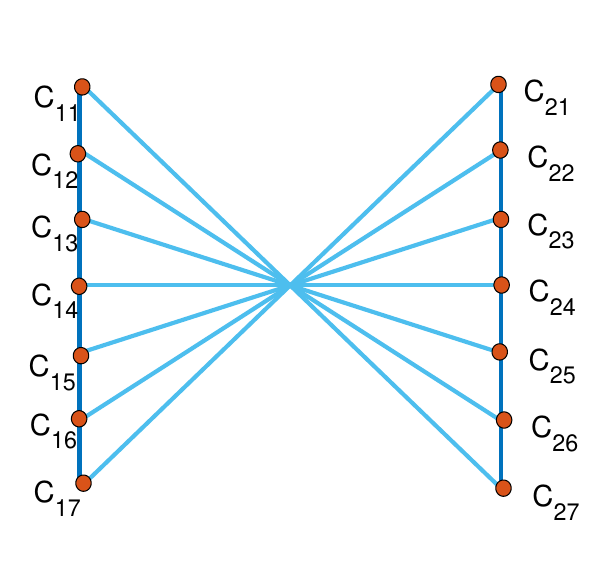}};
\node[anchor= west] (b)
  {\includegraphics[width=4cm]{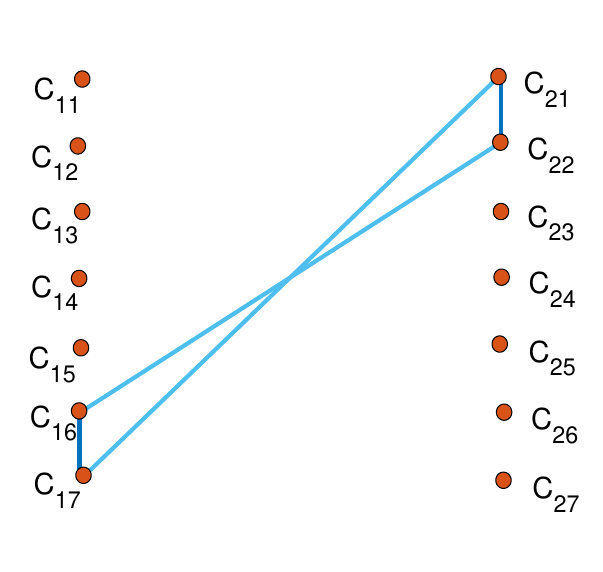}};
\endscope
\foreach \n in {a,b} {
  \node[anchor=west] at (\n.north west) {(\n)};
}

\end{tikzpicture}

\caption{\label{fig:2} The red dots symbolize the network nodes, each corresponding to a distinct type of lattice point within the three-dimensional Ising model. Meanwhile, the blue bonds illustrate the transformational relationships among these various nodes. (a) (b) represents two distinct and stable network configurations observed within the Ising model. c represents the weight of the node. }
	\label{fig2}
 
\end{figure}

In the preceding section, this article has elucidated the intricate correspondence between lattice models and network models. Next, I delve into the discussion of phase transitions through the utilization of the aforementioned network model, introducing a novel computational approach for analyzing such transitions.

\subsection{two-dimensional Ising model}

Firstly, at very low temperatures, the principle of detailed balance dictates that a portion of lattice points undergo a transition from $c_{15}$ to $c_{21}$. Concurrently, during this transition, the four adjacent lattice points surrounding each transforming point mutate into $c_{14}$. Given the extremely low temperature, instances where a lattice point is surrounded by other types of lattice points can be safely disregarded. Analogously, the principle of detailed balance also governs the transformation of some $c_{14}$ lattice points into $c_{22}$. During this conversion, additional $c_{14}$ lattice points emerge in the vicinity of the original $c_{14}$ points. Consequently, at lower temperatures, the weight distribution of lattice points becomes primarily concentrated on $c_{15}$, $c_{14}$, $c_{21}$, and $c_{22}$, resulting in a highly stable structural configuration as illustrated in Fig.~\ref{fig:1}(b). This stable configuration underscores the intricate interplay between lattice point transitions and the governing principles of detailed balance at low temperatures. In this paper, the network dominated by the aforementioned internal and boundary nodes is designated as a conservative component.

For the network presented in Fig.~\ref{fig:1}(a), when the weight of the central lattice point is initialized to unity, the central lattice point is constituted by the summation of $c_{23}$ and $c_{13}$, the distribution of lattice points across various stable configurations adheres to the principle of maximum entropy. This occurs because, during the continuous updating process, the central lattice point rapidly transforms into the weight of any other node, following the prescribed conversion formula. 

As the temperature escalates, the initial weight of $c_{15}$ is assigned a value of 1, thereby initiating the Ising model in the state depicted in Fig.~\ref{fig:1}(b). Upon traversing the phase transition point, the entire Ising model transitions to the state outlined in Fig.~\ref{fig:1}(a). 


In the preceding discussion, the model has been partitioned into the sum of two distinct components. One component, depicted in Fig.~\ref{fig:1}(b), primarily encompasses internal lattice points and boundary lattice $c_{14}$ points. Conversely, the second component is the maximum entropy model, illustrated in Fig.~\ref{fig:1}(a). The question then arises regarding how the weights of these two models evolve. Specifically, the inversion of a single voxel gives rise to the emergence of 2n boundary lattice points. Conversely, the inversion of boundary lattice points results in the generation of central lattice points.

For the two-dimensional Ising model, the inversion of an internal lattice $c_{15}$ point corresponds precisely to the generation of 8 central lattice points. To elaborate further, within Fig.~\ref{fig:1}, the transformation of a lattice point from $c_{15}$ to $c_{21}$ directly correlates with an augmentation of four $c_{15}$ lattice points. Subsequently, the inversion of $c_{14}$ lattice points gives rise to an increment of two in the population of both $c_{13}$ and $c_{23}$ lattice points. Consequently, the overall effect of the $c_{15}$ inversion is an alteration of 8 units in the collective quantity of $c_{13}$ and $c_{23}$ lattice points. However, for Ising models in dimensions exceeding two, a direct transformation between boundary lattice points and central lattice points does not exist. Instead, the concurrent inversion of multiple boundary lattice points is required to facilitate the emergence of a single central lattice point. By accounting for this intricate correspondence, we can derive the underlying relationships. In the context of the three-dimensional Ising model, the inversion corresponds to the emergence of three central lattice points, whereas for lattice points in four dimensions and beyond, it corresponds to the emergence of two central lattice points.

During the Monte Carlo simulation process for updating the Ising model, two distinct changes can be observed in the lattice points. Firstly, lattice points can undergo a direct spin flip with a certain probability, transforming from the node on the left side to the node on the right side, as depicted in Fig.~\ref{fig:1}(a) and (b). Secondly, a passive transformation occurs as a result of the first type of change, where the lattice point's type transitions to one of the two adjacent nodes on the same side. Due to the preservation of detailed balance, the weight associated with the first type of change is equivalent to twice the weight of the right-side node, denoted as $2p$. Consequently, the proportion of the component that does not directly participate in the spin flip is represented by $1-2p$. In the preceding discussion, it has been established that the flipping of an internal lattice point corresponds to the emergence of eight central lattice points. Precisely, during the process of flipping an internal lattice point, eight lattice points undergo passive transformations within the second type of change, and are subsequently selected with a probability denoted as $(1-2p)^8$.

Within the scope of this paper, the central lattice points are aligned with the maximum entropy model, reflecting the weights assigned to it. This alignment stems from the observation that when a particle is initially positioned at a central lattice point, the distribution of potential lattice points it can transform into adheres to the principles of the maximum entropy model. Put simply, as a lattice point transitions into a central lattice point, the subsequent transition from the central lattice point to other lattice point types follows the principle of maximum entropy. This fundamental principle underlies the formulation of the maximum entropy model.

Moreover, in Ising models with dimensions higher than two, boundary lattice points cannot undergo a direct transformation into central lattice points. Instead, there is an intermediate step involving the generation of other types of lattice points. Consequently, the lattice points located in the interfacial region, which serves as the dividing boundary between the two models, are composed of these diverse lattice point types. Collectively, these lattice points form the interfacial region that separates the internal lattice points from the central lattice points.


In Fig.~\ref{fig:1}, the lattice point $c_{14}$ and $c_{24}$ is designated as a boundary lattice point due to its analogous role to a boundary during the transition process. Subsequently, my focus shifts to exploring the transformational relationship between the boundary lattice points and the central lattice points. It is noteworthy that a phase transition occurs when the rate of change within the central lattice points surpasses that of the boundary lattice points.

As the weights shift from the boundary node $c_{14}$ towards $c_{13}$ and $c_{23}$, Monte Carlo simulations reveal that, upon selecting any lattice point associated with $c_{14}$, there exists a probability $2e^{-2/T}$ of its transformation into either $c_{13}$ or $c_{23}$. Additionally, for the boundary lattice point $c_{14}$, an increment in the number of lattice points results in a probability $1-e^{-4/T}$ of transitioning to $c_{22}$, as dictated by detailed balance principles. Consequently, as the temperature rises, the weights of the boundary lattice points initially exhibit an upward trend. However, once the ratio $2e^{-2/T}/(1-e^{-4/T})$ reaches unity, a reversal occurs, and the weights of the boundary lattice points commence a decline with further temperature elevations. As Illustrated in Fig.~\ref{fig:1}, we have identified two distinct stable configurations. Initially, as the temperature ascends, $c_{14}$ accumulates, resulting in the stabilization of the system in the configuration depicted in Fig.~\ref{fig:1}(b). Subsequently, as the weights of $c_{14}$ diminish and undergo rapid conversion into $c_{13}$ and $c_{23}$, the system transitions to the state portrayed in Fig.~\ref{fig:1}(a). By utilizing the aforementioned formula, we have successfully derived the rate equation governing the transformation of $c_{14}$. In Monte Carlo simulations, each reversal of a lattice point elicits corresponding modifications in its four neighboring lattice points, thereby resulting in a rate of change quantified as \begin{equation}
\label{eq:10}
   (\frac{2}{e^{2/T}-e^{-2/T}})^4.
\end{equation}
If the equation stated above evaluates to zero, it indicates the occurrence of a phase transition point.
At the initial stage, under extreme low-temperature conditions, for ferromagnetic systems, the spins in the Ising model predominantly align in the same direction. In quantitative terms, the weight associated with $c_{15}$ in Fig.~\ref{fig:1} approximates unity. As the temperature gradually increases, equilibrium is attained at various temperatures. Upon reaching the phase transition point, a state of maximum entropy and disorder is achieved, resembling the weight distribution depicted in Fig.~\ref{fig:1}(a). In this study, this evolution is conceptualized as a transition of weights from internal nodes to boundary nodes.

During the temperature rise, in Fig.~\ref{fig:1}(a), the probability of transition from the left node to the right node exhibits a steady increase. However, the transition from the left internal node to the boundary node in Fig.~\ref{fig:1}(a) may undergo an abrupt change at a critical temperature, signifying a shift in weight from internal to boundary nodes, and subsequently from boundary to central nodes.

Specifically, every time a spin-up lattice site is flipped in the model, a lattice site migrates from the left node to the right node, accompanied by the movement of four lattice sites to adjacent nodes. As illustrated in Fig.~\ref{fig:1}(b), when the weights of lattice sites are concentrated primarily on four nodes: $c_{15}$,$c_{14}$, $c_{21}$, and $c_{22}$, flipping a lattice site belonging to $c_{15}$ prompts the upward migration of these four unchanged lattice sites. Conversely, flipping a lattice site belonging to $c_{14}$ results in the upward migration of two unchanged lattice sites. Consequently, during the transition from $c_{15}$ to $c_{13}$ and $c_{23}$, eight unchanged lattice sites are involved, leading to a transition probability of $(1-2p)^8$.

From a conceptual standpoint, boundary lattice sites serve as intermediates in this transition. As the temperature rises, when a lattice site is added from $c_{15}$ to $c_{21}$, four boundary lattice sites are incorporated. These four lattice sites may transform into central lattice sites with a probability of $2e^{-2/T}$ or remain as boundary lattice sites with a probability of $1-e^{-4/T}$. The equilibrium relationship across various temperatures can be mathematically expressed as Eq.~\ref{eq:10}. Specifically, the evolution of eight lattice sites from the unchanged portion ultimately transforming into central lattice sites can be characterized by the probability of four boundary lattice sites in the unchanged portion remaining as boundary sites minus the probability of transforming into central sites. The ratio of this probability to the probability of four lattice sites remaining as boundary sites represents the probability of extracting eight lattice sites from the entire unchanged portion.

In probabilistic terms, since each transition involves eight lattice sites, they can be categorized into two distinct components: the unchanged portion and the changing portion. The unchanged portion corresponds to a weight of $1-2p$, while the changing portion corresponds to $2p$. If all eight randomly selected lattice sites belong solely to the unchanged portion, it reflects the probability of these sites remaining as boundary sites.

\begin{equation}
   1-(1-2p)^8=(\frac{2}{e^{2/T}-e^{-2/T}})^4
\end{equation}
Upon further analysis, it can be deduced that
\begin{equation}
    \langle m \rangle=1-2c=(1-\frac{1}{{sinh}^4(2/T)})^{1/8},T\le T_c 
\end{equation} 
\begin{equation}
    \langle m \rangle=0,T\ge T_c
\end{equation}

\subsection{three-dimensional Ising model}

\begin{figure}
\centering
\includegraphics[width=0.7\linewidth]{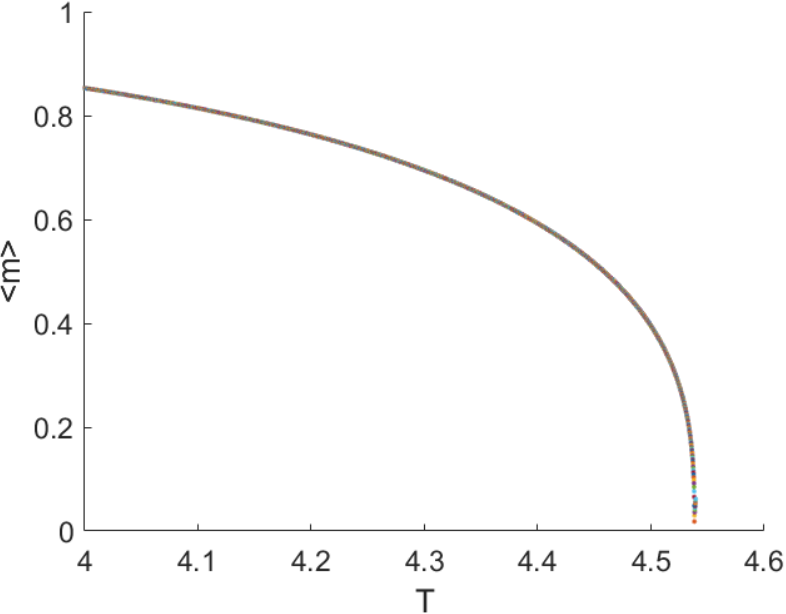}
\caption{\label{fig:3} The graph depicted above illustrates the variation of the average magnetic field strength with temperature in the three-dimensional Ising model.}
\end{figure}

The network structure has been decomposed into a maximum entropy component and a conservative component. In the transformed network, the transition probability of node weights exhibits a steady increase from left to right with rising temperatures. However, transitions among nodes within the same column may exhibit abrupt changes. This paper primarily focuses on the transitional relationships among nodes within a column. The competition between these two structures is effectively modeled as the interplay between internal nodes and central nodes through boundary nodes, encompassing mutual transformations between internal and boundary nodes, as well as between boundary and central nodes. From the perspective of nodal equilibrium, Monte Carlo updates continuously yield new boundary nodes, which concurrently transform into internal and central nodes.

In the context of the three-dimensional Ising model, the flipping of an internal lattice site corresponds to the emergence of six boundary lattice sites, and the simultaneous flipping of two boundary sites yields a central lattice site. Consequently, one internal lattice site is associated with three central lattice sites. For the four-dimensional and higher Ising models, the flipping of an internal lattice site generates 2n boundary lattice sites, and the simultaneous flipping of n-1 boundary sites is required to produce a central lattice site due to the integer constraint. This results in a one-to-two mapping between internal and central lattice sites. Applying this approach, we find that the exponent of $1-2p$ in the three-dimensional Ising model is 3, while in four-dimensional and higher models, it is 2.

The results obtained from the two-dimensional Ising model align precisely with the analytical solutions. Subsequently, our investigation progressed to the three-dimensional Ising model, as depicted in Fig.~\ref{fig:2}. In this three-dimensional configuration, each lattice point is flanked by six neighboring lattice points, implying that each spin flip affects the strengths of six bonds. Similarly, As the weights shift from the boundary node $c_{16}$ towards $c_{14}$ and $c_{24}$, Monte Carlo simulations reveal that, upon selecting any lattice point associated with $c_{16}$, there exists a probability $2e^{-4/T}$ of its transformation into either $c_{14}$ or $c_{24}$. Additionally, for the boundary lattice point $c_{16}$, an increment in the number of lattice points results in a probability $1-e^{-8/T}$ of transitioning to $c_{22}$, as dictated by detailed balance principles. Consequently, as the temperature rises, the weights of the boundary lattice points initially exhibit an upward trend. However, once the ratio $2e^{-4/T}/(1-e^{-8/T})$ reaches unity, a reversal occurs, and the weights of the boundary lattice points commence a decline with further temperature elevations. As Illustrated in Fig.~\ref{fig:2}, we have identified two distinct stable configurations. Initially, as the temperature ascends, $c_{16}$ accumulates, resulting in the stabilization of the system in the configuration depicted in Fig.~\ref{fig:2}(b). Subsequently, as the weights of $c_{16}$ diminish and undergo rapid conversion into $c_{14}$ and $c_{24}$, the system transitions to the state portrayed in Fig.~\ref{fig:2}(a). By utilizing the aforementioned formula, we have successfully derived the rate equation governing the transformation of $c_{16}$. In Monte Carlo simulations, each reversal of a lattice point elicits corresponding modifications in its six neighboring lattice points, thereby resulting in a rate of change quantified as 
\begin{equation}
   (\frac{2}{e^{4/T}-e^{-4/T}})^6
\end{equation}

we can also obtain

\begin{equation}
   1-(1-2p)^{3}=(\frac{2}{e^{4/T}-e^{-4/T}})^6  
\end{equation}

\begin{equation}
        \langle m \rangle=(1-\frac{1}{{sinh}^6(4/T)})^{1/3},T\le T_c 
\end{equation} 
\begin{equation}
    \langle m \rangle=0,T\ge T_c
\end{equation}
Utilizing the aforementioned formula, the critical exponent $\beta$ for the three-dimensional Ising model is determined as $1/3$.
Analogous to the two-dimensional case, we derived the following formula. Furthermore, we generalized this formula to encompass the Ising model in four-dimensional and higher models, Where n represents an integer$(n\ge 4)$.

\begin{equation}
   1-(1-2p)^{2}=(\frac{2}{e^{2(n-1)/T}-e^{-2(n-1)/T}})^{2n}
\end{equation}

\begin{equation}
        \langle m \rangle=(1-\frac{1}{{sinh}^{2n}(2(n-1)/T)})^{1/2},T\le T_c 
\end{equation} 
\begin{equation}
    \langle m \rangle=0,T\ge T_c
\end{equation}
Utilizing the aforementioned formula, the critical exponent $\beta$ for the high-dimensional Ising model is determined as 1/3.

The computational outcomes of the three-dimensional Ising model are presented in Fig.~\ref{fig:3}. To assess the accuracy of the derived formula, we compared the ratio of change rates between the two models with the Monte Carlo results obtained from the three-dimensional Ising model. The phase transition temperature of 4.511505\cite{butera2000extension,Blte1994SimulationsOT,kaupuvzs2001critical} for the three-dimensional Ising model, representing the phase transition point, is a widely accepted value within the scientific community.

\begin{equation}
   \mid 1-\frac{2}{e^{2(3-1)/4.511505}-e^{-2(3-1)/4.511505}} \mid
\end{equation}
Consequently, an error level of $0.7\%$ has been derived. The error originates from the presence of a small amount of $c_{15}$ in the model, as depicted in Fig.~\ref{fig:2}.
This comparison serves as a rigorous validation of my theoretical framework and enhances the scientific credibility of this findings.

Employing this methodology, the Ising model with an odd number of nearest-neighbor lattice points exhibits unique characteristics. Notably, the triangular lattice Ising model and the hexagonal lattice Ising model demonstrate properties that deviate significantly from those observed in the two-dimensional and three-dimensional Ising models. This deviation arises from the inability to directly translate the Ising model with odd nearest-neighbor lattice points into the summation of entropy networks, thus highlighting the intricacies and complexities inherent in these systems.

\subsection{xy model}

\begin{figure}
\centering
\includegraphics[width=0.7\linewidth]{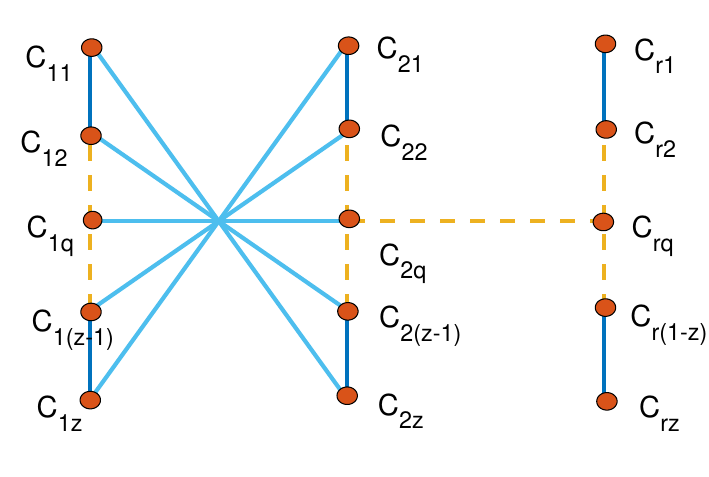}
\caption{\label{fig:4} The red dots symbolize the network nodes, each corresponding to a distinct type of lattice point within the three-dimensional Ising model. Meanwhile, the blue bonds illustrate the transformational relationships among these various nodes. c represents the weight of the node. "q" denotes the sequential index of the node attaining the average energy level, whereas "z" signifies the index of the node achieving the minimum energy state. The yellow dashed line serves to indicate the numerous nodes that are omitted from the depiction. Furthermore, "r" signifies the number of layers within the network, enumerated from left to right. }
\end{figure}

The xy model can be elegantly decomposed into the summation of two distinct yet stable network configurations. As depicted in Fig.~\ref{fig:4}, the phase transition within the xy model initiates with node $c_{1z}$ assigned a weight of 1, while all other nodes are assigned a weight of 0. Based on previous analytical insights, as the temperature escalates, a redistribution of weights occurs, flowing from $c_{1(z-1)}$ towards $c_{1q}$. Subsequently, once the weights reach $c_{1q}$, they propagate rapidly to $c_{2q}$, and so forth, ultimately culminating in a swift flow towards $c_{mq}$. Additionally, within each individual layer, the redistribution of weights is directed towards nodes exhibiting the highest and lowest energy levels, while bypassing those with average energy. This dynamic process gives rise to a structural configuration that adheres to the principle of maximum entropy, spanning from the second layer to the mth layer. This intricate structure serves as the foundation for the emergence of vortices within topological phases, thus providing a profound understanding of the underlying mechanisms governing the phase transition within the xy model.

\section{Conclusion}

To delve into the statistical laws governing lattice models and address the computational challenges associated with the exponential growth of complexity as the model scale increases, I transformed the lattice model into a network model that exhibits minimal sensitivity to model size. Initially, I categorized all potential lattice sites within the lattice model using the Hamiltonian, subsequently mapping these categories onto individual nodes in the network model. Then, leveraging Monte Carlo algorithms, I established relationships among the various nodes and correlated Monte Carlo updates with updates in the network model. Moreover, this approach aligns well with the findings of the renormalization group theory.

Subsequently, this study decomposes the constructed network model into a maximum entropy network structure and a conservative network counterpart. The investigation into the Ising and XY models offers profound insights into thermodynamic and KT phase transitions within the network model. The occurrence of phase transitions can be conceptualized as a competitive interplay between these two network configurations. Furthermore, employing the JQ model to delve into trivial quantum phase transitions within the network model presents an intuitive and compelling result.

Ultimately, under specified initial conditions, this work derive an accurate formula for the temperature-dependent magnetic induction intensity in the two-dimensional Ising model by estimating the equilibrium relationships attained by two distinct network structures across varying temperatures, leveraging the intricate interplay between nodes. Specifically, this paper analysis reveals a critical exponent $\beta$ of 1/3 for the three-dimensional Ising model, exhibiting a mere $0.7\%$ deviation from numerical findings. For higher-dimensional Ising models, the critical exponent $\beta$ remains consistent at 1/2. These outcomes are in close alignment with established research.

\textbf{\textit{Acknowledgments---}}
\label{acknowledgments}
This paper is supported by the National Natural Science Foundation of China-China Academy of Engineering
Physics(CAEP)Joint Fund NSAF(No. U2230402).

\bibliography{sample}

\nocite{*}


\end{document}